\documentstyle[12pt]{article}

\begin{document}
\begin{flushright}
\baselineskip=12pt
UPR-971-T \\
\end{flushright}

\begin{center}
\vglue 1.0cm
{\Large\bf Gauge Symmmetry and Supersymmetry Breaking by Discrete
Symmetry}
\vglue 2.0cm
{\Large
Tianjun Li~\footnote{E-mail: tli@bokchoy.hep.upenn.edu,
phone: (215) 573-5820, fax: (215) 898-2010.}}
\vglue 1cm
{Department of Physics and Astronomy, 
University of Pennsylvania, \\
Philadelphia, PA 19104 \\}
\end{center}
\vglue 1.0cm
\begin{abstract}
We study the principles of the
  gauge symmetry and supersymmetry breaking due to the
 local or global discrete symmetries on the extra space manifold.
We show that the gauge symmetry breaking by Wilson line
 is the special case of the discrete symmetry approach
where all the discrete symmetries are global and act freely
on the extra space manifold.
As applications, we discuss the $N=2$ supersymmetric
 $SO(10)$ and $E_8$ breaking on the space-time
$M^4\times A^2$ and $M^4\times D^2$, 
and point out that similarly one can study
any $N=2$ supersymmetric $SO(M)$ breaking.
We also comment on the one-loop effective potential,
 the possible questions and generalization.
\\[1ex]
\\[1ex]

\end{abstract}

\vspace{0.5cm}
\begin{flushleft}
\baselineskip=12pt
December 2001\\
\end{flushleft}
\newpage
\setcounter{page}{1}
\pagestyle{plain}
\baselineskip=14pt

\section{Introduction}
Grand Unified Theory (GUT) gives us an simple
 and elegant understanding of the quantum numbers of quarks and leptons,
and the success of gauge coupling unification in the Minimal
Supersymmetric
Standard Model strongly supports
 this idea. Right now, the Grand Unified Theory at high energy scale has
been widely accepted, however, there are some problems in GUT: 
 the grand unified
gauge symmetry breaking mechanism, the doublet-triplet splitting problem,
and
the proton decay problem, etc.

Recently, a new scenario proposed to address above questions in GUT has
 been discussed extensively~\cite{SBPL, JUNI, JUNII, HJLL}.
 The key point is that the supersymmetric GUT
model exists in 5 or higher dimensions and is broken down to the
4-dimensional 
$N=1$ supersymmetric Standard like Model for
 the zero modes due to the 
discrete symmetries in the neighborhoods of the branes
or on the extra space manifolds, which
become non-trivial constraints on the multiplets and gauge generators in 
GUT~\cite{JUNII}. 
The attractive models have been constructed explicitly, where
the supersymmetric 5-dimensional and 6-dimensional 
GUT models are broken down to
the 4-dimensional $N=1$
 supersymmetric $SU(3)\times SU(2) \times U(1)^{n-3}$
model, where $n$ is the rank of GUT group, through the 
compactification on various orbifolds and manifolds. 
The GUT gauge symmetry breaking and doublet-triplet
splitting problems have been solved neatly by the
discrete symmetry projections.
Other interesting phenomenology, like $\mu$ problems, gauge coupling
unifications, non-supersymmetric GUT, gauge-Higgs unification,
proton decay, etc, have also been discussed~\cite{SBPL, JUNI, JUNII,
HJLL}. 

In this paper, we discuss the gauge symmetry and supersymmetry 
breaking by the local or global discrete symmetries on the extra space
manifold in general. We prove the theorems for the
gauge symmetry on the fixed point, line or hypersurface, and in
the bulk for the zero modes. And we explain why the high dimensional
supersymmetry is broken down to the 4-dimensional
$N=1$ supersymmetry. These general results form 
the foundation of this approach,
which we would like to call it the discrete symmetry approach.
We also show that
the gauge symmetry breaking by Wilson line is the
special case of the discrete symmetry approach, where
all the discrete symmetries are global and act freely on
the extra space manifold.

In the previous discussions~\cite{SBPL, JUNII}, 
the $SO(10)$ is broken by 
the Klein discrete symmetries $Z_2\times Z_2$ on the extra
space manifold. So, we would like to discuss the
$SO(10)$ breaking by one cyclic discrete symmetry.
 In addition, although from the
phenomenological point of view $E_8$ is not very interesting,
we would like to discuss the $E_8$ breaking because of technical 
interesting.
In short, we study the $N=2$ supersymmetric $SO(10)$ breaking
and $E_8$ breaking on the space-time $M^4\times A^2$ and
$M^4\times D^2$, where
$M^4$ is the 4-dimensional Minkowski space-time, $A^2$ is
the 2-dimensional annulus, and $D^2$ is the 2-dimensional disc. 
We list all the constraints on constructing
 the 4-dimensional $N=1$
supersymmetric $SU(3)\times SU(2) \times U(1)^{n-3}$ model for the zero
modes,
in which $n$ is the rank, {\it i. e.}, $n=5$ and $8$ for $SO(10)$ and
$E_8$, respectively.
Moreover, we point out that on the space-time $M^4\times A^2$ and
$M^4\times D^2$,
we can break any $N=2$ supersymmetric $SO(M)$ models down to the
4-dimensional $N=1$
supersymmetric $SU(3)\times SU(2)\times U(1)^{n-3}$ models for the zero
modes
in which $n$ is the rank of the group $SO(M)$.
Furthermore, we comment on the one-loop
 effective potential, possible questions and
generalization in the discrete symmetry approach.

We would like to explain our convention. For simplicity, we define 
the $n\times n$ diagonal matrix as $(\alpha_1, \alpha_2, ..., \alpha_n)$,
for example,  we define the $3\times 3$ diagonal matrix as  
\begin{equation}
  (\alpha_1, \alpha_2, \alpha_3) \equiv \left(\begin{array}{ccc}
    \alpha_1 & 0 & 0  \\ 
    0 & \alpha_2 & 0 \\ 
    0 & 0 & \alpha_3 \\
  \end{array} \right)~.~\,
\end{equation}
In addition, suppose $G$ is a Lie group and $H$ is a subgoup of $G$.
In general, for $G=SU(N)$ and $G=SO(N)$, $H$ can be
the subgroup of $U(N)$ and $O(N)$, respectively.
We denote the commutant of $H$ in $G$ as $G/H$, {\it i. e.},
\begin{equation}
G/H\equiv \{g \in G|gh=hg, ~{\rm for ~any} ~ h \in H\}~.~\,
\end{equation}
And if $H_1$ and $H_2$ are the subgroups of $G$,
we define
\begin{equation}
G/\{H_1 \cup H_2\}\equiv \{G/H_1\} \cap \{G/H_2\}~.~\,
\end{equation}

\section{Gauge Symmmetry and Supersymmetry Breaking by Discrete Symmetry}
We assume that in a (4+n)-dimensional space-time manifold $M^4\times E$
where 
$M^4$ is the 4-dimensional Minkowski space-time and $E$ is the extra
n-dimensional space manifold, there exist
some local or global discrete symmetries on the extra space manifold and
the Lagrangian is invariant under the local or global discrete symmetries.
Moreover, the discrete symmetries may not act freely on $E$. 
 When the discrete symmetry 
does not act freely on $E$, there exists a brane at 
each fixed point, line or hypersurface, where the Standard Model fermions
can be located. 

Suppose we have $K$ discrete symmetries, and each discrete symmetry forms
a cyclic group $\Gamma_J$, for $J=1, 2, ...K$. Let us use the $I-th$ 
discrete symmetry as a representative for discussion, which
forms the cyclic group $\Gamma_I$.
As we know, the local discrete symmetry is defined in the
special brane's open neighborhood, where the position of the special brane
is
the only fixed point, line or hypersurface as long as the open
neighborhood
is small enough~\cite{JUNI, JUNII}.
 In the open neighborhood of the special brane,
the local discrete symmetry is global. So, if the $I-th$ discrete
 symmetry is local, we can concentrate on the
open neighborhood of the special brane. In short,
 without loss of generality, we can
 assume that the $I-th$ discrete symmetry is global for simplicity. 

Assume the coordinates for the extra dimensions are $y^1$, $y^2$,
..., $y^n$, the action of any element 
$\gamma_i^I$ $\subset$ $\Gamma_I$ on $E$ can be expressed as
\begin{eqnarray}
\gamma_i^I: ~~~( y^1, y^2, ..., y^n) \longrightarrow (\gamma_i^I y^1, 
\gamma_i^I y^2, ..., \gamma_i^I y^n) ~.~\,
\end{eqnarray}

The Lagrangian is invariant under the discrete symmetry, i. e., for any
element
$\gamma_i^I$ $\subset$ $\Gamma_I$
\begin{eqnarray}
{\cal L} (x^{\mu}, \gamma_i^I y^1, \gamma_i^I y^2, ..., \gamma_i^I y^n)
={\cal L} (x^{\mu},  y^1, y^2, ..., y^n) ~.~\,
\end{eqnarray}
So, for a generic bulk multiplet $\Phi$
which fills a representation of the bulk gauge group $G$, we have
\begin{eqnarray}
\Phi (x^{\mu}, \gamma_i^I y^1, \gamma_i^I y^2, ..., \gamma_i^I y^n) =
\eta_{\Phi}^I
(R_{\gamma_i^I})^{l_\Phi} \Phi (x^{\mu},  y^1, y^2, ..., y^n) 
(R_{\gamma_i^I}^{-1})^{m_\Phi}
~,~\,
\end{eqnarray} 
where $\eta^I_{\Phi}$ can be determined from the Lagrangian 
(up to $\Gamma_I$ for the matter fields), 
$\l_{\Phi}$ and $m_{\Phi}$ are the non-negative integers
determined by the representation of $\Phi$
under the bulk gauge group $G$.
In general, $\eta_{\Phi}^I$ is an element in $\Gamma_I$,
for example $\Gamma_I=Z_2$, $\eta_{\Phi}^I =\pm 1$. Moreover,
$R_{\gamma_i^I}$ is
an element in $G$, and $R_{\Gamma_I}$ is a discrete subgroup of $G$. We
will 
choose $R_{\gamma_i^I}$ as 
the matrix representation for $\gamma_i^I$ in the adjoint representation
of the
gauge group $G$. The consistent condition for $R_{\gamma_i^I}$ is
\begin{eqnarray}
R_{\gamma_i^I} R_{\gamma_j^I} = R_{\gamma_i^I \gamma_j^I} ~,~ 
\forall \gamma_i^I,~\gamma_j^I \subset \Gamma_I ~.~\,
\end{eqnarray} 
Mathematical speaking, the map 
$R:~ \Gamma_I \longrightarrow R_{\Gamma_I} \subset G $ is
a homomorphism. 

Furthermore, suppose $\Gamma_I=Z_N$, for a generic field $\phi$ in $\Phi$,
we have 
\begin{eqnarray}
\phi (x^{\mu}, \gamma_i^I y^1, \gamma_i^I y^2, ..., \gamma_i^I y^n) = 
\omega^{\kappa_i^I(\phi)} \phi (x^{\mu},  y^1, y^2, ..., y^n) 
~,~\,
\end{eqnarray}
where
\begin{eqnarray}
\omega=e^{{\rm i} {{2 \pi}\over N}} 
~,~\,
\end{eqnarray} 
and $\kappa_i^I(\phi)$ is an integer which can be determined from Eq. (6).
And if $\kappa_i^I(\phi)=0$ mod $N$, $\phi$ has zero mode.
\vglue 0.8cm
{\bf Theorem I.} For the zero modes, the gauge symmetry is
$G/\{R_{\Gamma_1} \cup R_{\Gamma_2} \cup ... \cup R_{\Gamma_K}\}$.
\vglue 0.8cm
{\bf Proof.} We first consider the $I-th$ discrete symmetry and assume
that
$\gamma_i^I$ is the generator of the cyclic group $\Gamma_I$. For the
gauge field $A=A_{\mu}^a T^a$, $\eta_A^I = 0$. So, for a fixed field 
$A_{\mu}^a T^a$, we have
\begin{eqnarray}
A_{\mu}^a (x^{\mu}, \gamma_i^I y^1, \gamma_i^I y^2, ..., \gamma_i^I y^n)
T^a = A_{\mu}^a (x^{\mu},  y^1,  y^2, ..., y^n)
R_{\gamma_i^I} T^a 
(R_{\gamma_i^I}^{-1})
~.~\,
\end{eqnarray} 
In addition, if $A_{\mu}^a T^a$ has zero mode, it satisfies the equation
\begin{eqnarray}
A_{\mu}^a (x^{\mu}, \gamma_i^I y^1, \gamma_i^I y^2, ..., \gamma_i^I y^n)
T^a = A_{\mu}^a (x^{\mu},  y^1,  y^2, ..., y^n) T^a 
~.~\,
\end{eqnarray} 
Then, we obtain that $A_{\mu}^a T^a$ has zero mode only if
$[R_{\gamma_i^I}, T^a]=0$, {\it i. e.}, $R_{\gamma_i^I}$ and $T^a$
are commute. So, under the discrete symmetry $\Gamma_I$, the
gauge symmetry is $G/R_{\Gamma_I}$ for the zero mode.
Because we have the same results for the other discrete symmetries, 
 for the zero modes, the gauge symmetry is
$G/\{R_{\Gamma_1} \cup R_{\Gamma_2} \cup ... \cup R_{\Gamma_K}\}$.
The theorem is proved.
\vglue 0.8cm
{\bf Theorem II.} If $(y^1=u^1, y^2=u^2, ..., y^n= u^n)$ is the 
fixed point of $\Gamma_I$ and is not the fixed point of
the rest discrete symmetries, the gauge symmetry
at that fixed point is $G/R_{\Gamma_I}$.
\vglue 0.8cm
{\bf Proof.} Suppose $\gamma_i^I$ is the generator of the cyclic group
$\Gamma_I$. At the fixed point $(y^1=u^1, y^2=u^2, ..., y^n= u^n)$
 of $\Gamma_I$, we have
\begin{eqnarray}
A_{\mu}^a (x^{\mu}, \gamma_i^I u^1, \gamma_i^I u^2, ..., \gamma_i^I u^n)
T^a
= A_{\mu}^a (x^{\mu},  u^1,  u^2, ..., u^n) T^a
~.~\,
\end{eqnarray}
Similar to the proof of Theorem I, we obtain that the gauge symmetry
at that fixed point is $G/R_{\Gamma_I}$. We would like to emphasize that
the gauge symmetry $G/R_{\Gamma_I}$ is preserved for all the KK modes.

\vglue 0.8cm
{\bf Corollary I.} Suppose $H$ is the subgroup of $\Gamma_I$, at
the fixed point of H which is not the
fixed point of $\Gamma_I$ and
the rest discrete symmetries, the gauge symmetry is 
$G/R_{H}$ where $R_{H}\subseteq R_{\Gamma_I}$.
\vglue 0.8cm

{\bf Corollary II.} If $\Gamma_I$ and $\Gamma_J$ have common
fixed point which is not the fixed point of
the rest discrete symmetries, the gauge symmetry at that 
common fixed point is 
$G/\{R_{\Gamma_I} \cup R_{\Gamma_J}\}$.
\vglue 0.8cm

{\bf Corollary III.} Suppose $S \subset E$ is the fixed hypersurface
of $\Gamma_I$ and is not fixed under the rest discrete symmetries,
 the gauge symmetry on $S$ is $G/R_{\Gamma_I}$.
In addition, if $S$ is the common fixed hypersurface
of $\Gamma_I$ and $\Gamma_J$,
and $S$ is not fixed under the rest discrete symmetries,
 the gauge symmetry on $S$ is 
$G/\{R_{\Gamma_I} \cup R_{\Gamma_J}\}$.
\vglue 0.8cm

{\bf Supersymmetry Breaking}. High dimensional supersymmetry
corresponds to the 4-dimensional $N>1$ supersymmetry. In terms of
the 4-dimensional $N=1$ supersymmetry language, the gauge 
multiplet can be decomposed into a vector
multiplet $V$ and the chiral multiplets $\Phi_1$, $\Phi_2$, ...,
$\Phi_{2k+1}$ in the adjoint representation.
Under the discrete symmetries, $\partial_{y^i}$ or 
$\partial_{z^i}$
in the complex coordinate may not be invariant,
for example, under the reflection $Z_2$ symmetry on the $i-th$ coordinate
$y^i$, $\partial_{y^i} \rightarrow - \partial_{y^i}$. 
Then, under the discrete symmetry, 
the transformations of the chiral mutlipets $\Phi_1$, $\Phi_2$, ...,
$\Phi_{2k+1}$ may be different from the transformation of the
vector multiplet $V$. Therefore, if all the component fields of
the chiral multiplets $\Phi_1$, $\Phi_2$, ...,
$\Phi_{2k+1}$ did not have the zero modes, we have the 4-dimensional $N=1$
supersymmetry for the zero modes. Similarly, one can discuss the
4-dimensional supersymmetry on the fixed point, lines or hypersuface.
By the way, we might keep the zero modes of
some fields in the chiral multiplets, which can be considered as
$SU(2)_L$ Higgs doublets in some models. 
\vglue 0.8cm
{\bf Relation to the Wilson Line Approach}. Assume that
$\Gamma_I$ is a global symmetry and acts freely on the extra
space manifold $E$, we can define the quotient manifold
$B=E/\Gamma_I$. Because $\Gamma_I$ belongs to the fundamental group of
$B$, {\it i. e.},
\begin{eqnarray}
\Gamma_I \subseteq \pi_1 (B) ~,~\,
\end{eqnarray} 
 the discrete symmetry approach become the Wilson line
approach for $\Gamma_I$~\cite{Wilson, YH}, {\it i. e.},
the gauge symmetry breaking by Wilson line. 

Furthermore, if all the discrete symmetries are global and
act freely on the extra
space manifold $E$, we can define the quotient manifold
$B=E/(\Pi_{J=1}^K \Gamma_J)$. And then, $\Pi_{J=1}^K \Gamma_J$ 
belong to the fundamental group of
$B$, {\it i. e.},
\begin{eqnarray}
\Pi_{J=1}^K \Gamma_J \subseteq \pi_1 (B) ~.~\,
\end{eqnarray} 
Therefore, the Wilson line approach is the special 
case of the discrete symmetry approach
where all the discrete symmetries are global and act freely
on the extra space manifold. 

To be explicit,
let us give an example. Consider the 5-dimensional
$N=1$ supersymmetric GUT models on 
the space-time $M^4\times S^1$, we can not break the 4-dimensional
$N=2$ supersymmetry down to $N=1$ supersymmetry in the Wilson
line approach. However, one can do this in the discrete
symmetry approach by considering the refelction symmetries ~\cite{SBPL,
JUNI}.
And if we moduloed the refelction symmetries, the fundamental group of
the extra space orbifold is trivial, {\it, i. e.},
$\pi_1(S^1/Z_2)$ or $\pi_1(S^1/(Z_2\times Z_2'))$ is trivial.

\section{$N=2$ Supersymmetric Theory on $M^4\times A^2$ and $M^4\times
D^2$}
In this section, we would like to review the $N=2$ supersymmetric theory
on the space-time $M^4\times A^2$ and $M^4\times D^2$, where $A^2$ and
$D^2$
are the 2-dimensional annulus and disc, respectively.

The convenient coordinates for the annulus $A^2$ 
is polar coordinates $(r, \theta)$,
and it is easy to change them to the complex 
coordinates by $z=r e^{{\rm i}\theta}$.
We call the innner radius of the annulus as $R_1$, and
the outer radius of the annulus as $R_2$. When $R_1=0$, 
the annulus becomes
 the disc $D^2$, which is an special case of $A^2$. 
We can define the $Z_n$ symmetry on the annulus $A^2$ by the
equivalent class
\begin{eqnarray}
z\sim \omega z~,~\,
\end{eqnarray} 
where $\omega=e^{{\rm i} {{2 \pi}\over n}}$. And we denote
the corresponding generator for $Z_n$ as $\Omega$ which satifies
$\Omega^n=1$. 
The KK mode expansions and the detail of this set-up can be
found in Ref.~\cite{JUNII}.

The $N=2$ supersymmetry in 6-dimension has 16 supercharges and
 corresponds to the $N=4$ supersymmetry in 4-dimension,
thus, only the gauge multiplet can be introduced in the bulk.  This
multiplet can be decomposed under the 4-dimensional
 $N=1$ supersymmetry into a vector
multiplet $V$ and three chiral multiplets $\Sigma$, $\Phi$, and $\Phi^c$
in the adjoint representation, with the fifth and sixth 
components of the gauge
field, $A_5$ and $A_6$, contained in the lowest component of $\Sigma$.
The Standard Model fermions are on the boundary 4-brane at $r=R_1$
or $r=R_2$ for the annulus $A^2$ scenario, 
and on the 3-brane at origin or on the boundary 4-brane at $r=R_2$ for 
the disc $D^2$ scenario. 

In the Wess-Zumino gauge and 4-dimensional $N=1$ supersymmetry
language, the bulk action 
is~\cite{NAHGW}
\begin{eqnarray}
  S &=& \int d^6 x \Biggl\{
  {\rm Tr} \Biggl[ \int d^2\theta \left( \frac{1}{4 k g^2} 
  {\cal W}^\alpha {\cal W}_\alpha + \frac{1}{k g^2} 
  \left( \Phi^c \partial \Phi   - \frac{1}{\sqrt{2}} \Sigma 
  [\Phi, \Phi^c] \right) \right) + {\rm h.c.} \Biggr] 
\nonumber\\
  && + \int d^4\theta \frac{1}{k g^2} {\rm Tr} \Biggl[ 
  (\sqrt{2} \partial^\dagger + \Sigma^\dagger) e^{-V} 
  (-\sqrt{2} \partial + \Sigma) e^{V}\Biggr]
\nonumber\\
&&+ \int d^4\theta \frac{1}{k g^2} {\rm Tr} \Biggl[
   \Phi^\dagger e^{-V} \Phi  e^{V}
  + {\Phi^c}^\dagger e^{-V} \Phi^c e^{V} 
\Biggr] \Biggr\}.
\label{eq:t2z6action}
\end{eqnarray}

From above action, we obtain  
the transformations of gauge multiplet under $\Omega$ as
\begin{eqnarray}
  V(\omega z, \omega^{n-1} \bar z) &=& (R_{\Omega})^{l_V}
 V(z, \bar z) (R_{\Omega}^{-1})^{m_V}~,~\,
\end{eqnarray}
\begin{eqnarray}
  \Sigma(\omega z, \omega^{n-1} \bar z) &=& \omega^{n-1}
(R_{\Omega})^{l_{\Sigma}} 
\Sigma(z, \bar z) (R_{\Omega}^{-1})^{m_{\Sigma}}~,~\,
\end{eqnarray}
\begin{eqnarray}
  \Phi(\omega z, \omega^{n-1} \bar z) &=& \omega^{n-1}
(R_{\Omega})^{l_{\Phi}} 
\Phi(z, \bar z) (R_{\Omega}^{-1})^{m_{\Phi}}~,~\,
\end{eqnarray}
\begin{eqnarray}
  \Phi^c(\omega z,  \omega^{n-1}\bar z) &=& \omega^{2}
(R_{\Omega})^{l_{\Phi^c}}  
\Phi^c(z, \bar z) (R_{\Omega}^{-1})^{m_{\Phi^c}}~,~\,
\end{eqnarray}
where $(l_V, m_V)$,  $(l_{\Sigma}, m_{\Sigma})$,
$(l_{\Phi}, m_{\Phi})$ and $(l_{\Phi^c}, m_{\Phi^c})$ are
determined from the representation under the
gauge group. Because we will decompose the bulk GUT group $G$
into its maximal subgroup, which is the product of several groups, 
for example 
$G\supset G_1\times G_2 \times G_k$, we define
$(l_V, m_V)$ as 
$(l_V^1, m_V^1)\otimes (l_V^2, m_V^2) \otimes 
... \otimes (l_V^k, m_V^k)$, where $(l_V^i, m_V^i)$
is determined from the decomposed representation of $V$ under
$G_i$. Similar notation will be used for
$(l_{\Sigma}, m_{\Sigma})$,
$(l_{\Phi}, m_{\Phi})$ and $(l_{\Phi^c}, m_{\Phi^c})$.  
To be explicit, we will give $(l_V, m_V)$,  $(l_{\Sigma}, m_{\Sigma})$,
$(l_{\Phi}, m_{\Phi})$ and $(l_{\Phi^c}, m_{\Phi^c})$
in detail in the following discussions.

\section{$SO(10)$ Breaking on $M^4\times A^2$ and $M^4\times D^2$}
In the previous discussions~\cite{SBPL, JUNII}, the $SO(10)$ is 
broken by the Klein discrete symmetries $Z_2\times Z_2$ on the extra
space manifold. So, we would like to discuss the
$SO(10)$ breaking by one cyclic discrete symmetry on the
space-time $M^4\times A^2$ and $M^4\times D^2$.
We will break the 6-dimensional $N=2$ supersymmetric 
$SO(10)$ model down to 4-dimensional $N=1$ supersymmetric
$SU(3)\times SU(2)\times U(1)^2$ model for the zero modes.
Because $SO(10) \supset SU(5)\times U(1)$, 
$SU(4)\times SU(2) \times SU(2)$ and $SO(8)\times U(1)$~\cite{Group},
and we can not obtain the Standard Model from
$SO(8)\times U(1)$, we decompose the $SO(10)$ 
into $SU(5)\times U(1)$ and $SU(4)\times SU(2) \times SU(2)$ 
in the following discussions of $SO(10)$ breaking. 

\subsection{Model I: $SO(10) \supset SU(5)\times U(1)$}
In this subsection, we discuss the $SO(10)$ breaking by 
considering its maximal subgroup 
$SU(5)\times U(1)$. 

The gauge fields of $SO(10)$ are in the adjoint representation 
of $SO(10)$ with dimension {\bf 45}. Under the gauge group
$SU(5)\times U(1)$, the $SO(10)$ gauge fields
decompose as~\cite{Group}
\begin{eqnarray}
{\bf 45= (24, 0) \oplus (10, 4) \oplus (\bar {10}, -4) \oplus
(0, 1)}~,~\,
\end{eqnarray} 

Assume that we have $Z_n$ symmetry on $A^2$ or $D^2$
and $\Omega$ is a generator of
$\Gamma=Z_n$, we choose the following matrix representation
for $\Omega$, which will give us the representations 
of all the elements in $\Gamma$,
\begin{equation}
R_{\Omega}=(\omega^{n_1}, \omega^{n_1}, \omega^{n_1},
\omega^{n_2}, \omega^{n_2}) \otimes (+1)~,~\,
\end{equation}
where $\omega=e^{{\rm i} {{2 \pi}\over n}}$. 
Therefore, we only need to concentrate on $SU(5)$.
In addition, $(l_V, m_V)$,  $(l_{\Sigma}, m_{\Sigma})$,
$(l_{\Phi}, m_{\Phi})$ and $(l_{\Phi^c}, m_{\Phi^c})$  
 are equal to $(1, 1)$ 
if the gauge field were
in the representation ${\bf (24, 0)}$, 
and $(l_V, m_V)$,  $(l_{\Sigma}, m_{\Sigma})$,
$(l_{\Phi}, m_{\Phi})$ and $(l_{\Phi^c}, m_{\Phi^c})$  
 are equal to $(2, 0)$ 
if the gauge fields were
in the representation ${\bf (10, 4)}$, 
and $(l_V, m_V)$,  $(l_{\Sigma}, m_{\Sigma})$,
$(l_{\Phi}, m_{\Phi})$ and $(l_{\Phi^c}, m_{\Phi^c})$  
 are equal to $(0, 2)$ 
if the gauge fields were
in the representation ${\bf ({\bar {10}}, -4)}$.
   
In order to have the models with $SU(3)\times SU(2) \times U(1)^2$
gauge symmetry and 4-dimensional $N=1$ supersymmetry 
for the zero modes,
we obtain the following constraints on $n_i$
\begin{eqnarray}
{\rm (a)~}  3 n_1+ 2 n_2  =0 ~{\rm mod}~n~,~\,
\end{eqnarray}
\begin{eqnarray}
{\rm (b)}~n_1\not= n_2 ~{\rm mod}~n~,~\,
\end{eqnarray} 
\begin{eqnarray}
{\rm (c)~} |n_1-n_2| \not= 1 ~{\rm and}~ n-2 ~{\rm mod}~n~,~\,
\end{eqnarray}
\begin{eqnarray}
{\rm (d)~} |n_i+n_j| \not= 0, 1 ~{\rm and}~ n-2 ~{\rm mod}~n~,~
~{\rm for}~i, j=1, 2 ~.~\,
\end{eqnarray}
Because $R_{\Omega} \subset
SU(5)\times U(1)$, we obtain the constraint
(a). And the constraint
 (b) will break the $SU(5)$ down to
$SU(3)\times SU(2)\times U(1)$. 
In addition, the constraint (c) will project out all the
zero modes of $\Sigma$, $\Phi$ and $\Phi^c$ in the 
representation ${\bf (24, 0)}$,
and the constraint (e) will project out all the zero modes of 
$V$, $\Sigma$, $\Phi$ and $\Phi^c$ in the representations 
${\bf (10, 4)}$ and ${\bf ({\bf \bar 10}, -4)}$. By the way, the
zero modes of $\Sigma$, $\Phi$ and $\Phi^c$ in the 
representation ${\bf (0, 1)}$ are automatically
projected out.

Let us give the model with $Z_{16}$ symmetry, 
the matrix representation for $R_{\Omega}$ is
\begin{eqnarray}
R_{\Omega} = (\omega^2, \omega^2, \omega^2, \omega^{5}, \omega^{5})
\otimes (+1)~.~\,
\end{eqnarray}
It is easy to check that all the constraints are satisfied.

First, we consider that the extra space manifold is the annulus $A^2$. 
For the zero modes, we have 4-dimensional $N=1$ supersymmetry and
$SU(3)\times SU(2) \times U(1)^2$ gauge symmetry in the bulk and on
the 4-branes at $r=R_1$ and $r=R_2$. Including the
KK states, we will have the 4-dimensional $N=4$ supersymmetry and
 $SO(10)$ gauge symmetry in the bulk, and on
the 4-branes at $r=R_1$ and $r=R_2$. 

Second, we consider that the extra space manifold is
the disc $D^2$. For the zero modes, 
we have 4-dimensional $N=1$ supersymmetry and
$SU(3)\times SU(2) \times U(1)^2$ gauge symmetry in the bulk and on
the 4-brane at $r=R_2$. Including all the
KK states, we will have the 4-dimensional $N=4$ supersymmetry and
 $SO(10)$ gauge symmetry in the bulk, and on
the 4-brane at $r=R_2$. In addition, because the origin 
($r=0$) is the fixed point under the $Z_{16}$ symmetry,
 we always have the 4-dimensional $N=1$ supersymmetry and
$SU(3)\times SU(2) \times U(1)^2$ gauge symmetry on the 3-brane at origin
in which only the zero modes exist.
 And if we put the Standard Model fermions on the 3-brane at origin,
 the extra dimensions
can be large and the gauge hierarchy problem can be solved 
for there does not exist the proton decay problem at all.

\subsection{Model II: $SO(10) \supset SU(4)\times SU(2)\times SU(2)$}
In this subsection, we discuss the $SO(10)$ breaking by 
considering its maximal subgroup 
$SU(4)\times SU(2) \times SU(2)$. 

The gauge fields of $SO(10)$ are in the adjoint representation 
of $SO(10)$ with dimension {\bf 45}. Under the gauge group
$SU(4)\times SU(2)_L \times SU(2)_R$, the $SO(10)$ gauge fields
decompose as~\cite{Group}
\begin{eqnarray}
{\bf 45=(15, 1, 1)\oplus (1, 3, 1)\oplus (1, 1, 3)\oplus(6, 2, 2)}~.~\,
\end{eqnarray}  

Assume that we have $Z_n$ symmetry on $A^2$ or $D^2$
and $\Omega$ is a generator of
$\Gamma=Z_n$, we choose the following matrix representation
for $\Omega$, which will give us the representations of 
all the elements in
$\Gamma$,
\begin{equation}
R_{\Omega}=(\omega^{n_1}, \omega^{n_1}, \omega^{n_1},
\omega^{n_2})
 \otimes (+1, +1)
\otimes (\omega^{n_3}, \omega^{n_4})~,~\,
\end{equation}
where $\omega=e^{{\rm i} {{2 \pi}\over n}}$. 
Moreover, $(l_V, m_V)$,  $(l_{\Sigma}, m_{\Sigma})$,
$(l_{\Phi}, m_{\Phi})$ and $(l_{\Phi^c}, m_{\Phi^c})$  
 are equal to $(1, 1)\otimes (0, 0) \otimes (0, 0)$ 
if the gauge fields were
in the representation ${\bf (15, 1, 1)}$, 
and $(l_V, m_V)$,  $(l_{\Sigma}, m_{\Sigma})$,
$(l_{\Phi}, m_{\Phi})$ and $(l_{\Phi^c}, m_{\Phi^c})$  
 are equal to $(0, 0)\otimes (1, 1) \otimes (0, 0)$ 
if the gauge fields were
in the representation ${\bf (1, 3, 1)}$, 
and $(l_V, m_V)$,  $(l_{\Sigma}, m_{\Sigma})$,
$(l_{\Phi}, m_{\Phi})$ and $(l_{\Phi^c}, m_{\Phi^c})$  
 are equal to $(0, 0)\otimes (0, 0) \otimes (1, 1)$ 
if the gauge fields were
in the representation ${\bf (1, 1, 3)}$, 
 and $(l_V, m_V)$,  $(l_{\Sigma}, m_{\Sigma})$,
$(l_{\Phi}, m_{\Phi})$ and $(l_{\Phi^c}, m_{\Phi^c})$
 are equal to
$(2, 0)\otimes (1, 0) \otimes (1, 0)$ if the gauge fields were
in the representation ${\bf (6, 2, 2)}$.
  
In order to have the models with $SU(3)\times SU(2) \times U(1)^2$
gauge symmetry and 4-dimensional $N=1$ supersymmetry 
for the zero modes,
we obtain the following constraints on $n_i$
\begin{eqnarray}
{\rm (a)~}  3 n_1+n_2  =0 ~{\rm mod}~n~,~\,
\end{eqnarray}
\begin{eqnarray}
{\rm (b)~}  n_3+n_4 =0 ~{\rm mod}~n~,~\,
\end{eqnarray}
\begin{eqnarray}
{\rm (c)}~n_1\not= n_2 ~{\rm mod}~n~,~\,
\end{eqnarray} 
\begin{eqnarray}
{\rm (d)}~n_3\not= n_4 ~{\rm mod}~n~,~\,
\end{eqnarray} 
\begin{eqnarray}
{\rm (e)~} |n_1-n_2| \not= 1 ~{\rm and}~ n-2 ~{\rm mod}~n~,~\,
\end{eqnarray}
\begin{eqnarray}
{\rm (f)~} |n_3-n_4| \not= 1 ~{\rm and}~ n-2 ~{\rm mod}~n~,~\,
\end{eqnarray}
\begin{eqnarray}
{\rm (g)~} n_i+n_j+n_k \not= 0, 1 ~{\rm and}~ n-2 ~{\rm mod}~n,
~{\rm for}~i, j=1, 2 ~{\rm and}~ k=3, 4~.~\,
\end{eqnarray}
Because $R_{\Omega} \subset
SU(4)\times SU(2)_L \times SU(2)_R$, we obtain the constraints
(a) and (b). And the constraints
 (c) and (d) will break the $SU(4)$ down to
$SU(3)\times U(1)$ and $SU(2)_R$ down to $U(1)$, respectively.
In addition, the constraints (e) and (f) will project out all the
zero modes of $\Sigma$, $\Phi$ and $\Phi^c$ in the 
representations ${\bf (15, 1, 1), (1, 1, 3)}$,
and the constraint (g) will project out all the zero modes of 
$V$, $\Sigma$, $\Phi$ and $\Phi^c$ in the representation
${\bf (6, 2, 2)}$. By the way, the
zero modes of $\Sigma$, $\Phi$ and $\Phi^c$ in the 
representation ${\bf (1, 3, 1)}$ are automatically
projected out.

Let us give the model with $Z_{16}$ symmetry, 
the matrix representation for $R_{\Omega}$ is
\begin{eqnarray}
R_{\Omega} = (\omega^{5}, \omega^{5}, \omega^{5}, \omega)
\otimes (+1, +1)
\otimes (\omega^{3}, \omega^{13})~.~\,
\end{eqnarray}
It is easy to check that all the constraints are satisfied.

First, we consider that the extra space manifold is the annulus $A^2$. 
For the zero modes, we have 4-dimensional $N=1$ supersymmetry and
$SU(3)\times SU(2) \times U(1)^2$ gauge symmetry in the bulk and on
the 4-branes at $r=R_1$ and $r=R_2$. Including the
KK states, we will have the 4-dimensional $N=4$ supersymmetry and
 $SO(10)$ gauge symmetry in the bulk, and on
the 4-branes at $r=R_1$ and $r=R_2$. 

Second, we consider that the extra space manifold is
the disc $D^2$. For the zero modes, 
we have 4-dimensional $N=1$ supersymmetry and
$SU(3)\times SU(2) \times U(1)^2$ gauge symmetry in the bulk and on
the 4-brane at $r=R_2$. Including all the
KK states, we will have the 4-dimensional $N=4$ supersymmetry and
 $SO(10)$ gauge symmetry in the bulk, and on
the 4-brane at $r=R_2$. Moreover, because the origin 
($r=0$) is the fixed point under the $Z_{16}$ symmetry,
 we always have the 4-dimensional $N=1$ supersymmetry and
$SU(3)\times SU(2) \times U(1)^2$ gauge symmetry on the 3-brane at origin
in which only the zero modes exist.

\subsection{$SO(M)$ Breaking on $M^4\times A^2$ and $M^4\times D^2$}
On the space-time $M^4\times A^2$ and $M^4\times D^2$, we can break
any 6-dimensional $N=2$ supersymmetric $SO(M)$ models down
to the 4-dimensional $N=1$ supersymmetric 
$SU(3)\times SU(2) \times U(1)^{n-3}$ models where $n$ is the
rank of the group $SO(M)$. The method is similar to above. We first
decompose
the $SO(M)$ group into the product of $SU(m_i)$ and/or $U(1)$ groups,
then project out all the zero modes of the non-Standard Model like gauge 
fields and all chiral multiplets.
 For instance, $SO(12)$, we can decompose it into
$SU(6)\times U(1)$ or $SU(4)\times SU(4)$, then discuss $SO(12)$ breaking
by using above method and the branching rules for various representations 
in Ref.~\cite{Group}.

\section{$E_8$ Breaking on $M^4\times A^2$ and $M^4\times D^2$}
Although from the phenomenological point of 
view $E_8$ is not interesting, we will discuss the $E_8$ breaking on the
space-time $M^4\times A^2$ and $M^4\times D^2$ because of
technical interesting.  
We will show how to break the 6-dimensional $N=2$ supersymmetric 
$E_8$ model down to 4-dimensional $N=1$ supersymmetric
$SU(3)\times SU(2)\times U(1)^4$ model for the zero modes.
By the way, we would like to point out that
on the 6-dimensional space-time 
$M^4\times S^1/(Z_2\times Z_2') \times S^1/(Z_2\times Z_2')$,
we can break the 6-dimensional $N=2$ supersymmetric 
$E_8$ model down to 4-dimensional $N=1$ supersymmetric
 $SO(10)\times U(1)$ model, or 
$SU(3)\times SU(3)\times SU(3)\times SU(2)\times U(1)$ model, etc,
however, we can not break it down to the 4-dimensional $N=1$
supersymmetric $SU(3)\times SU(2)\times U(1)^4$ model for the zero modes
 unless we consider the 8-dimensional space-time
$M^4\times S^1/(Z_2\times Z_2') \times S^1/(Z_2\times Z_2')
\times S^1/(Z_2\times Z_2') \times S^1/(Z_2\times Z_2')$.

$E_8$ has following maximal subgroups:
$SU(3)\times E_6$, $SU(4)\times SO(10)$, $SU(5)\times SU(5)$,
$SU(2)\times E_7$, $SO(16)$ and $SU(9)$~\cite{Group}. And the first three
maximal subgroups are interesting in the weakly coupled heterotic 
$E_8\times E_8$ string theory or 
M-theory on $S^1/Z_2$ model buildings, so, we decompose the
gauge fields according to the first three maximal subgroups. As we know,
the gauge fields of $E_8$ are in the adjoint representation
with dimension {\bf 248}, 
under the maximal subgroup $SU(3)\times E_6$, the gauge
fields decompose as~\cite{Group}
\begin{eqnarray}
{\bf 248=(8, 1) \oplus (1, 78) \oplus (3, 27) \oplus (\bar 3, \bar
{27})}~,~\,
\end{eqnarray}
and under the maximal subgroup $SU(4)\times SO(10)$, the gauge
fields decompose as
\begin{eqnarray}
{\bf 248=(15, 1) \oplus (1, 45) \oplus (4, 16) \oplus (\bar 4, \bar {16})
\oplus (6, 10)}~,~\,
\end{eqnarray}
and under the maximal subgroup $SU(5)\times SU(5)$, the gauge
fields decompose as~\cite{Group}
\begin{eqnarray}
{\bf 248=(24, 1) \oplus (1, 24) \oplus (5, 10) \oplus (\bar 5, \bar {10})
\oplus (10, \bar 5) \oplus (\bar {10}, 5)}~.~\,
\end{eqnarray}
In this paper, we only discuss the $E_8$ breaking by decomposing
$E_8$ into $SU(5)\times SU(5)$. Similarly, noticing that the
$E_6$ breaking has been discussed in Ref.~\cite{HJLL} and
$SO(10)$ breaking has been discussed in the last section, and using
the branching rules for various representations in Ref.~\cite{Group},
one can discuss the $E_8$ breaking by decomposing $E_8$ into
$SU(3)\times E_6$ or $SU(4)\times SO(10)$.

Assume that we have $Z_{2n}$ symmetry on $A^2$ or $D^2$
and $\Omega$ is a generator of
$\Gamma=Z_{2n}$, we choose the following matrix representation
for $\Omega$, which will give us the representations of 
all the elements in $\Gamma$,
\begin{equation}
R_{\Omega}=(+1, +1, +1, \omega^n, \omega^n)
\otimes (\omega^{n_1}, \omega^{n_2}, \omega^{n_3},
\omega^{n_4}, \omega^{n_5})
 ~,~\,
\end{equation}
where $\omega=e^{{\rm i} {{2 \pi}\over {2n}}}$. 
Moreover, $(l_V, m_V)$,  $(l_{\Sigma}, m_{\Sigma})$,
$(l_{\Phi}, m_{\Phi})$ and $(l_{\Phi^c}, m_{\Phi^c})$  
 are equal to $(1, 1)\otimes (0, 0)$ 
if the gauge fields were
in the representation ${\bf (24, 1)}$, 
and $(l_V, m_V)$,  $(l_{\Sigma}, m_{\Sigma})$,
$(l_{\Phi}, m_{\Phi})$ and $(l_{\Phi^c}, m_{\Phi^c})$  
 are equal to $(0, 0)\otimes (1, 1)$ 
if the gauge fields were
in the representation ${\bf (1, 24)}$, 
and $(l_V, m_V)$,  $(l_{\Sigma}, m_{\Sigma})$,
$(l_{\Phi}, m_{\Phi})$ and $(l_{\Phi^c}, m_{\Phi^c})$  
 are equal to $(1, 0)\otimes (2, 0)$ 
if the gauge fields were
in the representation ${\bf (5, 10)}$, 
 and $(l_V, m_V)$,  $(l_{\Sigma}, m_{\Sigma})$,
$(l_{\Phi}, m_{\Phi})$ and $(l_{\Phi^c}, m_{\Phi^c})$
 are equal to
$(0, 1)\otimes (0, 2)$ if the gauge fields were
in the representation ${\bf (\bar 5, \bar {10})}$,
and $(l_V, m_V)$,  $(l_{\Sigma}, m_{\Sigma})$,
$(l_{\Phi}, m_{\Phi})$ and $(l_{\Phi^c}, m_{\Phi^c})$  
 are equal to $(2, 0)\otimes (0, 1)$ 
if the gauge fields were
in the representation ${\bf (10, \bar 5)}$, 
 and $(l_V, m_V)$,  $(l_{\Sigma}, m_{\Sigma})$,
$(l_{\Phi}, m_{\Phi})$ and $(l_{\Phi^c}, m_{\Phi^c})$
 are equal to
$(0, 2)\otimes (1, 0)$ if the gauge fields were
in the representation ${\bf (\bar {10}, 5)}$,
  
In order to have the models with $SU(3)\times SU(2) \times U(1)^4$
gauge symmetry and 4-dimensional $N=1$ supersymmetry 
for the zero modes,
we obtain the following constraints on $n_i$
\begin{eqnarray}
{\rm (a)~}  n_1+n_2 +n_3+n_4+n_5 =0 ~{\rm mod}~2n~,~\,
\end{eqnarray}
\begin{eqnarray}
{\rm (b)}~n_i\not= n_j ~{\rm mod}~2n~,
~{\rm for}~i, j=1, 2, 3, 4, 5 ~{\rm and}~i\not= j~,~\,
\end{eqnarray} 
\begin{eqnarray}
{\rm (c)~} |n_i-n_j| \not= 1 ~{\rm and}~ n-2 ~{\rm mod}~2n~,
~{\rm for}~i, j=1, 2, 3, 4, 5 ~{\rm and}~i\not= j~,~\,
\end{eqnarray}
\begin{eqnarray}
{\rm (d)~} |n_i| \not= 0, 1 ~{\rm and}~ n-2 ~{\rm mod}~2n~,
~{\rm for}~i=1, 2, 3, 4, 5~,~\,
\end{eqnarray}
\begin{eqnarray}
{\rm (e)~} |n-n_i| \not= 0, 1 ~{\rm and}~ n-2 ~{\rm mod}~2n~,
~{\rm for}~i=1, 2, 3, 4, 5~,~\,
\end{eqnarray}
\begin{eqnarray}
{\rm (f)~} |n_i+n_j| \not= 0, 1 ~{\rm and}~ n-2 ~{\rm mod}~2n~,
~{\rm for}~i, j=1, 2, 3, 4, 5~,~\,
\end{eqnarray}
\begin{eqnarray}
{\rm (g)~} |n_i+n_j+n| \not= 0, 1 ~{\rm and}~ n-2 ~{\rm mod}~2n~,
~{\rm for}~i, j=1, 2, 3, 4, 5~.~\,
\end{eqnarray}
Because $R_{\Omega} \subset
SU(5)\times SU(5)$, we obtain the constraint
(a). And the constraint
 (b) will break the second $SU(5)$ down to
$U(1)^4$.
In addition, the constraint (c) will project out all the
zero modes of $\Sigma$, $\Phi$ and $\Phi^c$ in the 
representation ${\bf (1, 24)}$,
and the constraints (d), (e), (f) and (g)
 will project out all the zero modes of 
$V$, $\Sigma$, $\Phi$ and $\Phi^c$ in the representations 
${\bf (5, 10)}$, ${\bf (\bar 5, \bar {10})}$,
${\bf (10, \bar 5)}$ and ${\bf (\bar {10}, 5)}$.
By the way, the
zero modes of $\Sigma$, $\Phi$ and $\Phi^c$ in the 
representation ${\bf (24, 1)}$ are automatically
projected out if $n > 2$.

Let us give the model with $Z_{66}$ symmetry, 
the matrix representation for $R_{\Omega}$ is
\begin{eqnarray}
R_{\Omega} = (+1, +1, +1, -1, -1)\otimes
(\omega^{3}, \omega^{6}, \omega^{9}, \omega^{12}, \omega^{36})~.~\,
\end{eqnarray}
It is easy to check that all the constraints are satisfied.

First, we consider that the extra space manifold is the annulus $A^2$. 
For the zero modes, we have 4-dimensional $N=1$ supersymmetry and
$SU(3)\times SU(2) \times U(1)^4$ gauge symmetry in the bulk and on
the 4-branes at $r=R_1$ and $r=R_2$. Including the
KK states, we will have the 4-dimensional $N=4$ supersymmetry and
 $E_8$ gauge symmetry in the bulk, and on
the 4-branes at $r=R_1$ and $r=R_2$. 

Second, we consider that the extra space manifold is
the disc $D^2$. For the zero modes, 
we have 4-dimensional $N=1$ supersymmetry and
$SU(3)\times SU(2) \times U(1)^4$ gauge symmetry in the bulk and on
the 4-brane at $r=R_2$. Including all the
KK states, we will have the 4-dimensional $N=4$ supersymmetry and
 $E_8$ gauge symmetry in the bulk, and on
the 4-brane at $r=R_2$. In addition, because the origin 
($r=0$) is the fixed point under the $Z_{66}$ symmetry,
 we always have the 4-dimensional $N=1$ supersymmetry and
$SU(3)\times SU(2) \times U(1)^4$ gauge symmetry on the 3-brane at origin
in which only the zero modes exist.
 And if we put the Standard Model fermions on the 3-brane at origin,
 the extra dimensions
can be large and the gauge hierarchy problem can be solved 
for there does not exist the proton decay problem at all.

\section{Remarks, Possible Questions and Generalization}
First, we would like to comment on the one-loop effective potential.
It was pointed out that the representations ($R_{\Gamma_I}$)
 of the discrete symmetries ($\Gamma_I$) are not
 arbitrary parameters of the theory and are determined 
dynamically~\cite{YH}. 
The tree-level effective potential is trivial, so, we need to 
consider the one-loop effective potential which can be done by the
well-known results of the background field gauge. However, if there
exist non-broken supersymmetry at the compactification scale, it is
not difficult for us to show that the one-loop contribution is zero.
Therefore, the 4-dimensional $N=1$ 
supersymmetry breaking scale must be much lower than
the compactification scale, which is true for the supersymmetric 
GUT models in Ref.~\cite{SBPL, JUNI, JUNII, HJLL} where the 4-dimensional
$N=1$
supersymmetry breaking scale is around TeV, and
the compactification scale or GUT scale is around $10^{16}$ GeV.
This also implies that if we considered the non-supersymmetric GUT
breaking
by discrete symmetry, we must calculate the one-loop effective
potential and determine the representations ($R_{\Gamma_I}$)
 of the discrete symmetries ($\Gamma_I$) dynamically, {\it i. e.},
we can not choose $R_{\Gamma_I}$ arbitrarily.

However, there might still exist the question in the discrete
symmetry approach. 
Because the vacua are degenerate, why the nature chooses the
Standard Model like gauge symmetry as vaccum. For example, we consider
the supersymmetric $SU(5)$ model, the $SU(5)$ gauge symmetry can
be unbroken, or broken down to the $SU(4)\times U(1)$ or
$SU(3)\times SU(2) \times U(1)$ gauge symmetry, 
the question is why the $SU(5)$ Model
is broken down to the Standard Model.

Second, when we consider the 6-dimensional supersymmetric GUT breaking,
we often consider the 6-dimensional $N=2$ supersymmetry in order to
avoid the gauge anomaly. Because we can not have the hypermultiplets in
the bulk, we have to put the Standard Model fermions
and Higgs doublets on the fixed point or line. So, we need to
understand how to localize the matter fields and Higgs fields at the
fixed point or line. 

Third, we can not reduce the rank of GUT group
in the discrete symmetry approach because the cyclic group is abelian.
If we want to reduce the rank of the GUT group, we might need to 
consider the continuous symmetry on the extra space manifold. 
It is also interesting to generalize the
discrete symmetry approach on
the non-commutative extra dimensions.

\section{Conclusion}
We study the principles of the
  gauge symmetry and supersymmetry breaking due to the
 local or global discrete symmetries on the extra space manifold.
We show that the gauge symmetry breaking by Wilson line
 is the special case of the discrete symmetry approach
where all the discrete symmetries are global and act freely
on the extra space manifold.
As applications, we discuss the $N=2$ supersymmetric
 $SO(10)$ and $E_8$ breaking on the space-time
$M^4\times A^2$ and $M^4\times D^2$, 
and point out that similarly one can study
any $N=2$ supersymmetric $SO(M)$ breaking.
We also comment on the one-loop effective potential,
 the possible questions and generalization.

\section*{Acknowledgments}
This work was supported in part 
 by the U.S.~Department of Energy under Grant 
 No.~DOE-EY-76-02-3071.

\end{document}